\documentclass[pra,superscriptaddress,showpacs,papersize=a4paper,twocolumn,floatfix]
{revtex4-1}%
\usepackage[pdftex]{graphicx}
\usepackage{bm}
\usepackage{subcaption}
\usepackage{etoolbox}
\usepackage[usenames,dvipsnames]{color}
\usepackage{mathtools}
\usepackage{amsmath}%
\usepackage{amsfonts}%
\usepackage{amssymb}

\DeclarePairedDelimiterX\braket[2]{\langle}{\rangle}{#1 \delimsize\vert #2}

\newcommand{\bg}{ \begin{gather} }

\newcommand{\eg}{\end{gather}}
\newcommand{\be}{ \begin{equation} }
\newcommand{\ee}{\end{equation}}
\newcommand{\bea}{ \begin{eqnarray} }
\newcommand{\eea}{\end{eqnarray}}

\def\eps{\epsilon}
\def\tr{\textrm{tr}}

\newcommand{\sgn}{\mathop{\rm sgn}}
\renewcommand{\Re}{\mathop{\rm Re}}

\DeclareMathOperator{\Sch}{Sch}
\DeclareMathOperator{\PSL}{PSL}
\newcommand{\RR}{\mathbb{R}}
\newcommand{\const}{\mathrm{const}}
\newcommand{\eff}{\mathrm{eff}}
\newcommand{\SYK}{\mathrm{SYK}}

\newcommand{\spf}{\mathrm{sp4}}
\newcommand{\spt}{\mathrm{sp2}}
\newcommand{\fluc}{\mathrm{fluc}}
\newcommand{\glass}{\mathrm{gl}}
\newcommand{\tbeta}{{\tilde{\beta}}}

\AtEndEnvironment{thebibliography}{
}

\begin{document}

\title{Perturbed Sachdev-Ye-Kitaev model: a polaron in the hyperbolic plane.}
\author{A. V. Lunkin}
\affiliation{Skolkovo Institute of Science and Technology, 143026 Skolkovo, Russia}
\affiliation{ L. D. Landau Institute for Theoretical Physics, Kosygin Str.~2, Moscow 119334, Russia}
\affiliation{Physics Department, National Research University "Higher School of Economics", Moscow, Russia}
\author{A. Yu.\ Kitaev}
\affiliation{California Institute of Technology, Pasadena, CA 91125, USA.}
\author{M. V. Feigel'man}
\affiliation{ L. D. Landau Institute for Theoretical Physics, Kosygin Str.~2, Moscow 119334, Russia}
\affiliation{Skolkovo Institute of Science and Technology, 143026 Skolkovo, Russia}

\begin{abstract}
	We study the SYK$_4$ model with a weak SYK$_2$ term of magnitude $\Gamma$ beyond the simplest perturbative limit considered previously. For intermediate values of the perturbation strength, $J/N \ll \Gamma \ll J/\sqrt{N}$, fluctuations of the Schwarzian mode are suppressed, and the SYK$_4$ mean-field solution remains valid beyond the timescale $t_0 \sim N/J$ up to $t_* \sim J/\Gamma^2$. Out-of-time-order correlation function displays at short time intervals exponential growth 
	with maximal Lyapunov exponent $2\pi T$, but its prefactor scales as $T$ at low temperatures $T \leq \Gamma$.
	
\end{abstract}

\date{\today}

\maketitle


\textit{Introduction}.  The Sachdev-Ye-Kitaev model~\cite{Kitaev2015,Kitaev2018soft,maldacena2016remarks,bagrets2016sachdev} presents a rare example of an analytically tractable theory of strongly correlated fermions. While the original SYK$_4$ model with only fourth-order terms in the Hamiltonian looks exotic from the condensed matter perspective, the inclusion of quadratic terms brings the theory closer to the world of strongly interacting electrons. Thus, the SYK$_4$ Hamiltonian with a small SYK$_2$ perturbation is a reasonable model in this regard. In the mean-field approximation (which is exact in the limit of infinite number of Majorana modes $N$), the SYK$_2$ term is always dominant at temperature $T=0$~\cite{Altman,Davison,Yao,song2017strongly,Ref2,Ref5}. More exactly, the saddle point solution for the SYK$_4$ Green function, $G_{\spf}(t) \propto 1/\sqrt{t}$, turns to $G_{\spt}(t) \propto 1/t$ (characteristic of a Fermi liqid) at longer times. At finite $N$, the SYK$_4$ phase is stable to quadraic perturbations below some threshold~\cite{Lunkin2018} due to soft mode fluctuations, which change the asymptotic behavior of the Green function to $G_{\fluc}(t) \propto 1/t^{3/2}$ at $t \gg t_0 \sim N/J$~\cite{bagrets2016sachdev}. The stability result, originally obtained via straightforward perturbation theory~\cite{Lunkin2018}, was later generalized to several related models (tunnelling between different SYK grains, complex fermions) and to the renormalization group (RG) framework~\cite{BAK19-1,BAK19-2}. The threshold value of perturbation strength $\Gamma$ was estimated as $\Gamma_c \sim J/N$ and is more accurately determined below. If $\Gamma>\Gamma_c$, then the system behaves as a Fermi liquid at the longest times.

In the present Letter, we extend the results~\cite{Lunkin2018,BAK19-1,BAK19-2} in two important directions.  First, we present a method to analyze the SYK$_4$+SYK$_2$ model beyond perturbation theory or its RG-like variant.  Second, our new method works at a finite temperature (which is assumed to be greater than the temperature $T_{\glass}$ of an expected~\cite{SYKglass} glass transition). In particular, we study a previously unexplored range of intermediate strengths of the SYK$_2$ term, $J/N \ll \Gamma \ll J/\sqrt{N}$. In this regime, the quadratic perturbation is weak at the time scale $t \leq t_0 \sim N/J$, where the saddle-point (conformal) solution $G_{\spf}$ is applicable.  Paradoxically, we find that the perturbation \underline{stabilizes} the conformal solution $G_{\spf}(t) \sim 1/\sqrt{t}$ for extended times, $t \gg t_0$, where the Green function of the pure SYK$_4$ model is modified by the soft mode fluctuations.  Only at the longest time scale, $t \ge J/\Gamma^2 \gg t_0$, the conformal solution $G_{\spf}$ gives way to the Fermi-liquid solution $G_{\spt}(t) \propto 1/t$.

Recent experiments on strongly correlated electron systems~\cite{mablg1,mablg2} with nearly flat band
demonstrate a "strange metal" behaviour with resistivity $R(T) \propto T$  which was shown~\cite{song2017strongly} (within purely saddle-point analysis) to occur in an array of SYK "dots" coupled by weak tunneling. We demonstrate in this Letter that such a behaviour can be realized under 
much broader range of conditions than those implied in Ref.~\cite{song2017strongly}.

Our results are best undersood using the geometric interpretation~\cite{kitaev2019statistical} of the Schwarzian theory in terms of an auxiliary particle whose trajectories are closed curves in the hyperbolic plane. The quadratic perturbation is then described as the particle being coupled to a free scalar Bose field. For sufficiently strong coupling, $\Gamma\gg J/N$, a polaron-type bound state is formed, resulting in increased rigidity of the curve and the suppression of its fluctuations.


\textit{The model and basic equations.}
The Hamiltonian involves $N$ Majorana operators $\chi_1,\dots,\chi_N$ and has the following form:
\begin{equation}
H=\frac{1}{4!}\sum_{i,j,k,l} J_{ijkl}\chi_i\chi_j\chi_k\chi_l+\frac{i}{2!}\sum \Gamma_{ij}\chi_i\chi_j.
\label{eq:Hamiltonian}
\end{equation}
Here $(J_{ijkl})$ and $(\Gamma_{ij})$ are anti-symmetric tensors. Each of their components is a Gaussian random variable with zero mean and the following variance: $\overline{J_{ijkl}^2}= \frac{3! J^2}{N^3}$ and $\overline{\Gamma_{ij}^2}=\frac{\Gamma^2}{N}$. We assume that $\Gamma\ll J$. When averaging over disorder (i.e.\ over $J_{ijkl}$ and $\Gamma_{ij}$), the Green function is taken to be diagonal in replicas. This is self-consistent above the glass transition temperature $T_{\glass}$, where $\sqrt{N} \leq \ln(J/T_{\glass}) \leq \frac23 s_0 N$ with $s_0 = 0.4648..$, see~\cite{glass_comment}. We consider the problem in imaginary time, introduce new fields $G(\tau,\tau^\prime)$, and impose the constraint $G(\tau,\tau^\prime)=- (1/N) \sum_{i=1}^N\chi_i(\tau)\chi_i(\tau^\prime)$ using Lagrange multiplier fields $\Sigma(\tau,\tau')$. By performing the functional integral over the Grassmann variables $\chi_i$, we obtain an effective action for $G$ and $\Sigma$. The result has the form $S=S_{\SYK}+S_{2}$, where
\begin{eqnarray}
\frac{2}{N}S_{\SYK} &=& 
\begin{aligned}[t]
&\tr\ln(-\partial_\tau -\Sigma) \\
&+\int \left[ \Sigma(\tau,\tau^\prime)G(\tau,\tau^\prime)
-\frac{J^2}{4}G(\tau,\tau^\prime)^4 \right] d\tau\, d\tau^\prime
\end{aligned}
\label{eq:S_SYK}\\
\frac{2}{N}S_{2} &=&
-\frac{\Gamma^2}{2} \int G(\tau,\tau^\prime)^2\,d\tau\, d\tau^\prime.
\label{eq:S_2}
\end{eqnarray}
The action $S_{\SYK}$ \textit{with the derivative term neglected} --- which is applicable for $t\gg 1/J$ --- is invariant under reparametrizations of time. In particular, this abridged action has a degenerate saddle: along with a unique conformal (i.e.\ $\PSL(2,\RR)$-invariant) saddle point $(G_c,\Sigma_c)$, there is a manifold of related points $(G_\varphi,\Sigma_\varphi)$. It is parametrized by functions $\varphi$ that map the circle of length $\beta=1/T$ to the circle of length $2\pi$:
\begin{equation}
\begin{gathered}\label{Gphi}
G_{\varphi}(\tau_1,\tau_2)
=J^{-2\Delta}G_c\bigl(\varphi(\tau_1),\varphi(\tau_2)\bigr)
\varphi'(\tau_1)^{\Delta}\varphi'(\tau_2)^{\Delta},
\\
G_c(\varphi_1,\varphi_2)
=-b^{\Delta}\sgn(\varphi_{12})|\varphi_{12}|^{-2\Delta},
\end{gathered}
\end{equation}
where $\Delta=\frac{1}{4}$,\, $b=\frac{1}{4\pi}$, and $\varphi_{12}=2\sin\frac{\varphi_1-\varphi_2}{2}$. (The corresponding expressions for $\Sigma$ differ by the 	replacement $\Delta\to 1-\Delta$ and an overall factor of $J^2$.) The term with $\partial_\tau$ in the original action (\ref{eq:S_SYK}) lifts the degeneracy and leads to the ``Schwarzian action'' defined on the soft mode manifold~(\ref{Gphi}). Thus, the complete action (together with the quadratic term) becomes
\begin{equation}
\label{action0}
\begin{aligned}
S[\varphi]
={}& -\frac{\gamma}{J} \int_0^\beta  \Sch(e^{i\varphi(\tau)},\tau)\, d\tau\\
&-\frac{N\Gamma^2}{4}\int G_{\varphi}(\tau_1,\tau_2)^2\, d\tau_1\, d\tau_2
\end{aligned}
\end{equation}
where $\Sch(f(x),x)=\bigl(\frac{f''}{f'}\bigr)' -\frac{1}{2}\bigl(\frac{f''}{f'}\bigr)^2$ is the Schwarzian derivative, $\gamma=\alpha_S N$, and $\alpha_S\sim 0.01$ is a numerical constant~\cite{maldacena2016remarks,Kitaev2018soft}; also see a comment on that matter in~\cite{Lunkin2018}.

The use of action in the form of Eq.~(\ref{action0}) needs some explanation. Although the second term in that equation is equal to $S_2$ defined by~(\ref{eq:S_2}), the function $G_{\varphi}$ in it belongs to the soft mode manifold. In contrast, the $G$ in the original action $S_{\SYK}+S_{2}$ is completely general, so $S_2$ plays a dual role: i) it competes with the main term $S_{SYK}$ in determining the saddle point solution, and ii) it reduces soft mode fluctuations since its very presence breaks down the reparametrization symmetry. The strength of the first effect is controlled by the parameter $\Gamma/J$ and does not depend on $N$. This is why for small values of $\Gamma/J$, it is natural to start our analysis from effect ii), which is captured by Eq.~(\ref{action0}). We will see that the second term suppresses soft-mode fluctuations and extends the domain where the SYK$_4$ conformal solution is valid.


\textit{Geometric interpretation and the polaron analogy.} The following picture is conceptually important, but we will do most calculations by a different method. Therefore, this section will be brief; more details can be found in the Supporting Material.
The duality between the Schwarzian action and JT gravity was discussed in Ref.~\cite{kitaev2019statistical}
as well as in the papers~\cite{JT1,maldacena2016conformal,engelsoy2016investigation}

The geometric interpretation of the Schwarzian action~\cite{kitaev2019statistical} is based on a correspondence between functions $\varphi$ as described above and closed curves on the hyperbolic plane. Such curves may be parametrized by the proper length $\ell=J\tau$, which will be used instead of $\tau$ for the purpose of this discussion. In the Poincare disk model with metric $ds^2=\frac{4}{(1-r^2)^2}(dr^2+r^2 d\varphi^2)$, the curve is given by the equations $\varphi=\varphi(\ell)$ and $r=1-\varphi'(\ell)$. This representation is valid if $\varphi''(\ell)\ll \varphi'(\ell)\ll 1$, which is true for a typical curve of length $L=J\beta\gg 1$ in the statistical ensemble. Under the same conditions, we have $\Sch(e^{i\varphi(\ell)},\ell)= K-1$, where $K$ is the extrinsic curvature of the curve at the given point. This allows for an elegant representation of the Schwarzian action $S_{\Sch} =-\gamma\int_0^L \Sch(e^{i\varphi(\ell)},\ell)\, d\ell$ in terms of the length of the curve and the enclosed area; however, some regularization is necessary in order to define the functional integral~\cite{kitaev2019statistical}. Replacing the function $\varphi$ with the curve $X$, we may rewrite Eq.~(\ref{action0}) as follows:
\begin{equation}
\label{SX0}
S[X]=S_{\Sch}[X]-\frac{N\Gamma^2}{4J^2}\int G^2_{\Phi}\bigl(X(\ell_1),X(\ell_2)\bigr)\,
d\ell_1\, d\ell_2,
\end{equation} 
where $G_{\Phi}(r_1,\varphi_1;r_2,\varphi_2)\propto |\varphi_{12}|^{-2\Delta} (1-r_1)^{\Delta}(1-r_2)^{\Delta}$ near the disk boundary. The function $G_{\Phi}$ can be identified  with the propagator of a scalar boson $\Phi$. Thus, the nonlocal interaction between different points of the curve is decoupled, such that the action~(\ref{SX0}) is obtained from
\begin{equation}
\label{SXPhi}
S[X,\Phi]=S_{\Sch}[X]+S_{\Phi}[\Phi]+\int_{0}^{L} \Phi(X(\ell))\,d\ell 
\end{equation}
where $S_{\Phi}= \frac{1}{4g\gamma}\int d\mu \Phi(x)(-\nabla^2-\frac{1}{4}+\delta^2)\Phi(x)$,
by integrating out $\Phi$.

The action~(\ref{SXPhi}) is similar to the polaron problem, where an electron in a crystal interacts with an elastic deformation. By analogy with the heavy polaron, we will look for a mean-field solution where the field $\Phi$ forms a potential well close to the boundary of the Poincare disk. The general form of $\Phi$ in this region is $\Phi(r,\varphi) =\Lambda(\varphi)(1-r)^{\Delta}$, and the solution in question is $\Lambda(\varphi)=\const$. The curve roughly follows the circle $r= 1-\frac{2\pi}{L}$ and slightly wiggles. This behavior may be understood as a localized state of a quantum particle, whose coordinate is conveniently defined as $\xi=-\ln(\gamma(1-r))$.

\textit{Adiabatic action.}
We proceed with a formal solution for the polaron. It is convenient to rescale time as $\tau\to \frac{J\tau}{\gamma}$ and to introduce a similarly rescaled inverse temperature $\tbeta$ and a new coupling constant $g$:
\begin{equation}
\tbeta=\frac{J \beta}{\gamma},\qquad
g=\frac{b^{2\Delta}}{2}\frac{N\Gamma^2}{J^2}\gamma^{2-4\Delta}
= \frac{N\gamma}{4\sqrt{\pi}}\frac{\Gamma^2}{J^2}.
\label{g}
\end{equation}
Then the action (\ref{action0}) reads: 
\begin{equation}
S[\varphi] =
- \int_0^\tbeta  \Sch(e^{i\varphi(\tau)},\tau)\, d\tau
-\frac{g}{2}\int \left(\frac{\varphi_1^\prime  \varphi_2^\prime}{\varphi^2_{12}}\right)^{1/2}\, d\tau_1\, d\tau_2.
\label{Sphi2}
\end{equation}
Now we reduce the path integral with this action to some solvable quantum mechanical problem.  To implement this idea, we introduce new time-dependent variables $\xi(\tau) = -\ln\left(\varphi^\prime(\tau)\right)$ and $\Xi(\tau) = [\varphi'(\tau)]^{1/2}$, and the corresponding Lagrange multipliers $\lambda(\tau)$ and $\Lambda(\tau)$. This means inserting $\delta(\varphi'-e^{-\xi}) =\int_{-i\infty}^{+i\infty} \exp\bigl(\lambda(\varphi'-e^{-\xi})\bigr)\frac{d\lambda}{2\pi i}$ and $\delta(\Xi-e^{-\xi/2})$ (expressed likewise using $\Lambda$) in the functional integral. Thus, the action takes the form
\begin{widetext}
	\begin{eqnarray}
	S[\varphi,\xi,\lambda,\Xi,\Lambda]=\int_0^\tbeta \left(\frac{\xi^{\prime 2}}{2}-\lambda\left(\varphi^\prime-e^{-\xi}\right)-\frac{1}{2}e^{-2\xi}-\Lambda\left(\Xi-e^{-\xi/2}\right)\right)d\tau  
	-\frac{g}{2}\iint \frac{\Xi(\tau_1) \Xi_2(\tau_2) }{|\varphi_{12}|}\,d\tau_1\,d\tau_2.
	\label{Sphi3}
	\end{eqnarray} 
\end{widetext}
We assume that $\tbeta\gg 1$ so that the term $\frac{1}{2}e^{-2\xi}$ is relatively small. It will be neglected in our analysis.

We treat action (\ref{Sphi3}) using adiabatic approximation, with $\xi$ being the fast variable. That is, the functional integral of $e^{-S}$ over $\xi$ is performed under the assumption that $\varphi'(\tau)$, $\lambda(\tau)$, $\Xi(\tau)$, and $\Lambda(\tau)$ are constant at a suitable time scale $\tau_*$ (to be determined later). The result has the form $e^{-S_{\eff}}$, where 
\begin{equation}\label{Sground}
\begin{aligned}
S_{\eff}[\varphi,\lambda,\Xi,\Lambda] ={}& \int_0^{\tbeta}
\bigl(E_0(\lambda,\Lambda)-\lambda\,\varphi^\prime-\Lambda\,\Xi\,\bigr)\,d\tau\\
&- \frac{g}{2}\int |\varphi_{12}|^{-1} \Xi(\tau_1) \Xi(\tau_2)\, d\tau_1\, d\tau_2
\end{aligned}
\end{equation}
and $E_0(\lambda,\Lambda)$ is the ground state of the effective Hamiltonian for the variable $\xi$,
\begin{eqnarray}
\hat{H}_{\lambda,\Lambda}=-\frac{1}{2}\partial_{\xi}^2+\Lambda e^{-\xi/2}+\lambda e^{-\xi}.
\label{Hxi}
\end{eqnarray}
This Hamiltonian has bound states with energies 
\begin{equation}
E_n = -\frac{(\kappa-1-2n)^2}{32}, \qquad
n=0,\ldots, \left\lfloor \frac{\kappa-1}{2} \right\rfloor,
\label{En}
\end{equation}
where $\kappa =- \sqrt{\frac{8}{\lambda}}\Lambda$. The corresponding eigenfunctions $\psi_n(\xi)$ are provided in the Supporting Material. The characteristic time for the adiabatic approximation can be estimated as $\tau_* \sim(E_1-E_0)^{-1} =\frac{8}{\kappa-2}$. Such an estimate is certainly correct for a harmonic oscillator, where the oscillation period is the only relevant time scale. The Hamiltonian~(\ref{Hxi}) is similar if $\kappa\gg 1$. We will see that the last condition actually guarantees adiabaticity, i.e.\ that $\phi'$, $\lambda$, $\Xi$, $\Lambda$ do not fluctuate at the time scale $\tau_*$. In fact, the fluctuations at all time scales are small enough to be considered Gaussian.

Our next goal is to derive an effective action for $\varphi$. To this end, we find the saddle point of the action (\ref{Sground}) with respect to the other variables. The saddle point conditions for $\lambda$, $\Lambda$, and $\Xi$ read:
\begin{eqnarray}
&\varphi^\prime  = \frac{\partial E_0}{\partial\lambda}
=\frac{\kappa-1}{32}\frac{\kappa}{\lambda}\,, \qquad
\Xi = \frac{\partial E_0}{\partial\Lambda}
= -\frac{\kappa-1}{16}\frac{\kappa}{\Lambda}\,,
\label{Xiphi}\\
&\Lambda(\tau_1) = -g \int d\tau_2\, \frac{\Xi(\tau_2)}{|\varphi_{12}|}\,.
\label{Lambda}
\end{eqnarray}
Eqs.~(\ref{Xiphi}) allow one to eliminate $\lambda$ and $\Lambda$ from various formulas; in particular, the definition of $\kappa$ is equivalent to the relation $\Xi^2 = \frac{\kappa-1}{\kappa} \varphi^\prime$. The integrand in the first term of the action~(\ref{Sground}) can be written as
\begin{equation}
E_0(\lambda,\Lambda)-\lambda\,\varphi^\prime-\Lambda\,\Xi
=\frac{\kappa-1}{32},
\end{equation}
and Eq.~(\ref{Lambda}) becomes an equation for $\kappa(\tau)$:
\begin{eqnarray}
\label{kappa1}
\kappa^2(\tau_1)\,\eta(\tau_1) =
16g \int d\tau_2\,
\frac{\eta(\tau_2)\sqrt{\varphi'(\tau_1)\varphi'(\tau_2)}}{|\varphi_{12}|},
\end{eqnarray}
where $\eta(\tau) = \sqrt{1 - \kappa^{-1}(\tau)}$. Finally, the effective action is reduced to
\begin{equation}\label{Seff0}
S=\int_0^\tbeta \frac{\kappa-1}{32}d\tau
-\frac{g}{2}\int d\tau_1d\tau_2
\frac{\eta(\tau_1)\eta(\tau_2)\sqrt{\varphi'(\tau_1)\varphi'(\tau_2)}}
{|\varphi_{12}|}.
\end{equation}

Now, let $\kappa\gg1$ so that $\eta(\tau)\approx1$. Furthermore, we will assume (and later verify) that the fluctuations are small, and hence, both $\varphi^\prime\approx 2\pi/\tbeta$ and $\kappa$ are nearly constant. Then Eq.~(\ref{kappa1}) is simplified as follows:  
\begin{equation}
\kappa^2 = 	16g \int  \frac{d\varphi(\tau_2)}{\Bigl|2\sin\Bigl(\frac{\varphi(\tau_1)-\varphi(\tau_2)}{2}\Bigr)\Bigr|}
\approx 32 g \ln\biggl(\frac{\kappa\tbeta}{16\pi}\biggr), 
\label{eq:kappa_self_consistently}
\end{equation}
where we have used the cutoff $|\tau_1-\tau_2|>\tau_* \approx\frac{8}{\kappa}$ for the logarithmic integral. As for the effective action~(\ref{Seff0}), its first term may be neglected (see Supporting Material). Expressing $\varphi'$ as a function of $\varphi$, namely, $\varphi'(\tau)=\varepsilon(\varphi)$, we get:
\begin{equation}
S\approx -\frac{g}{2}
\int_{0}^{2\pi}\int_{0}^{2\pi}
\frac{d\varphi_1}{\varepsilon(\varphi_1)}
\frac{d\varphi_2}{\varepsilon(\varphi_2)}
\left(\frac{\varepsilon(\varphi_1)\varepsilon(\varphi_2)}
{\varphi_{12}^2}\right)^{1/2}.
\label{Seff1}
\end{equation}


\textit{Saddle point solution.}
The action~(\ref{Seff1}) attains its minimum at the constant field configuration, $\varepsilon(\varphi)=2\pi/\tbeta$, as well as all configurations related to it by $\PSL(2,\RR)$ symmetries. The minimum value of the action, $S_{\min}\approx -\tbeta\kappa^2/32$, determines a correction to the SYK free energy: $F\approx E_0 - Ns_0T - \frac{J}{\gamma}\frac{\kappa^2}{32}$.
Differentiating it and using Eq.(\ref{eq:kappa_self_consistently}), we find the entropy of the system:
\begin{equation}
\label{entropy}
S(T) \approx Ns_0-\frac{gJ}{\gamma T}
\end{equation}
Note that the entropy vanishes at $T\sim \Gamma^2/J$, which is roughly the temperature at which the $\SYK_4$ conformal Green function $G_{\spf}(\tau,0)\sim -(J\tau)^{-1/2}$ gives way to the Fermi liquid behavior, $G_{\spt}(\tau,0)\sim -(\Gamma\tau)^{-1}$.  At lower temperatures, $T \leq \Gamma^2/J$, we need to modify our polaron solution.  Namely, we should introduce an upper cutoff in Eq.~(\ref{eq:kappa_self_consistently}), $|\tau_1-\tau_2|< J/\Gamma^2$. Thus, the equation for $\kappa$ becomes
\begin{equation}\label{kappa_sc1}
\kappa^2=32g\ln\biggl(\frac{\kappa\overline{\beta}}{16\pi}\biggr),\qquad
\overline{\beta}=\frac{J}{\gamma}\min\biggl(\beta,\frac{J}{\Gamma^2}\biggr).
\end{equation}

\textit{Fluctuations.} Let us estimate the fluctuation of $\varepsilon(\varphi)$ around $\varepsilon_0=2\pi/\tbeta$ and show that they are small. For simplicity, we assume that $T\gg \Gamma^2/J$. In addition to the adiabatic action $S$ given by Eq.~(\ref{Seff1}), we need the first non-adiabatic correction. The latter is identical to the Schwarzian action (see Supplement). We consider the Fourier series $\varepsilon(\varphi) = \varepsilon_0 + \frac{1}{2\pi}\sum_{n} \delta\varepsilon_n e^{in\varphi}$, expand the effective action $S_{\Sch}+S$ up to the second order in $\delta\varepsilon_n$ with $n\not=0$, and calculate the Gaussian expectation values $\langle\delta\varepsilon_n\delta\varepsilon_{-n}\rangle$. (Note that $\delta\varepsilon_0$ is determined by the equation $\int \frac{d\varphi}{\varepsilon(\varphi)}=\tbeta$.) This calculation, which can be found in the Supplement, gives the following result:
\begin{equation}
K_\varepsilon(n) \equiv\frac{\langle\delta\varepsilon_n\delta\varepsilon_{-n}\rangle}{\varepsilon_0^2} =
\frac{2\pi\varepsilon_0}{\varepsilon_0^2(n^2-1) + (g/2)\tilde\psi(n)}
\label{eps-fluct}
\end{equation}
where $\tilde\psi(n) = \psi(n+\frac12)-\psi(-\frac12)$ and $\psi(x)$ is the digamma function; thus, $\tilde\psi(n) \approx \ln(n)$ for $n \gg 1$. Eq.~(\ref{eps-fluct}) is accurate for $n\ll n_{*} \equiv (\varepsilon_0\tau_*)^{-1} =\kappa\tbeta/(16\pi)$ because it was derived from the effective action that is valid at sufficiently long times, $\tau\gg \tau_* = 8/\kappa$. However, no inconsistency occurs at greater values of $n$: for $n\gtrsim n_*$, the first term in the denominator of~(\ref{eps-fluct}) starts to dominate over the second one, and the fluctuations are suppressed. The summation of the r.h.s.\ of Eq.~(\ref{eps-fluct}) over all $n$, or just those with $|n| \leq n_*$, leads to the estimate
\begin{equation}
\frac{\langle(\delta\varepsilon)^2\rangle}{\varepsilon_0^2} \approx 
\frac{1}{\epsilon_0 n_*} = \frac{8}{\kappa} \ll 1. 
\label{delta-eps}
\end{equation}
Thus, the fluctuations of $\varepsilon(\varphi)$ are much smaller than its typical value if $\kappa \gg 8$.

\begin{figure}
	\includegraphics{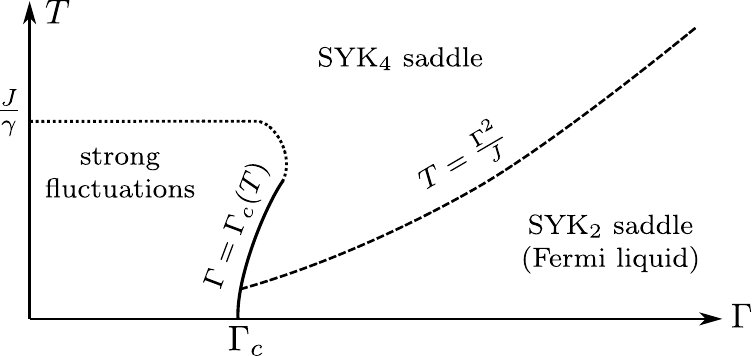}
	\caption{Regions with different Green function behaviors at large $\tau$. 
		The phase boundaries are not sharp, except for the boundary between the
		fluctuation region and saddle point regions (the solid line). The latter
		is well-defined asymptotically, under the condition  $T \ll J/\gamma$. Hence, the
		termination point of the solid line, given by the condition $ T \sim
		J/\gamma$, is fuzzy.}
	\label{fig:ph_diag}
\end{figure}

\textit{Phase diagram.}
As parameter $\kappa$ decreases toward unity, the fluctuations become strong and the adiabatic approximation breaks down. At $\kappa \sim 1$, we expect a transition into another phase of our model, where the $\SYK_2$ term is irrelevant at all time scales~\cite{Lunkin2018}.  In terms of the original parameters of the model, the transition occurs when $\Gamma$ becomes smaller than its critical value given by the equations 
\begin{equation}
\Gamma_c \sim \frac{J}{\sqrt{N\gamma\ln\frac{\overline{\beta}}{16\pi}}},\qquad
\overline{\beta}
=\frac{J}{\gamma}\min\biggl(\frac{1}{T},\frac{J}{\Gamma_c^2}\biggr),
\label{Gammac}
\end{equation}
where $\gamma=\alpha_SN$. Note that the critical value $\Gamma_c=\Gamma_c(T)$ decreases logarithmically with the decrease of the physical temperature $T$ while $T\gtrsim \frac{\Gamma_c^2(T)}{J}$. At lower temperatures, $\Gamma_c(T)$ remains constant. The lines $\Gamma=\Gamma_c(T)$ and $T=J/\gamma$ (see Figure~\ref{fig:ph_diag}) separate the region with strong fluctuations, characterized by the Green function $G_{\fluc}(\tau,0)\sim \gamma(J\tau)^{-3/2}$ for sufficiently large $\tau$ (but still much less than $\beta$), from regions where the saddle point solution is valid.

\textit{Higher-order Green functions.}
Conformal solution (\ref{Gphi}) describes single-particle fermionic Green function  of the original 
problem with Hamiltonian (\ref{eq:Hamiltonian}). Additional information on its quantum dynamics  is provided 
by higher-order fermion Green functions defined as $G^{(p)}(\tau,\tau^\prime) \equiv
\left( -\frac1{N}\sum_i\chi_i(\tau)\chi_i(\tau^\prime)\right)^p$.
It can be shown (see Supplementary Material, sec.III)
that  the functions $G^{(p)}(\tau,\tau^\prime) $ with  $ p \ll \kappa$ can be calculated by means of the effective action 
(\ref{Seff1}) and its propagator  (\ref{eps-fluct}).
To find them, we need just to average $p$-power of the  conformal solution (\ref{Gphi}) over fluctuations
of variables $\xi$ and $\varphi$ described by the polaron bound-state:  
$G^{(p)}(\tau_1,\tau_2)= (-1)^p\langle\left[b e^{-\xi_1-\xi_2}\sin^{-2}(\frac12(\varphi_1-\varphi_2))\right]^{p/4}\rangle$.
The result of calculations (provided in the SM, sec.III) reads (remember that $\kappa \ll N$):
\begin{equation}
\frac{G^{(p)}(\tau_1 - \tau_2)}{[G(\tau_1-\tau_2)]^p} = 
\exp\left[\frac{p^2}{4\kappa}\left(1 + f(\theta_{12})\right)\right]
\label{moments}
\end{equation}
where $\theta_{12} = 2\pi T (\tau_1-\tau_2)$ and function $f(\theta)$ is provided below ($n_*\varepsilon_0=\kappa/8$):
\begin{widetext}
	\begin{equation}
	f(\theta) = \frac{2+n_{*}\theta}{(n_{*}\theta)^{2}}\left[2n_{*}\theta\cosh\left(\frac{n_{*}\theta}{2}\right)-4\sinh\left(\frac{n_{*}\theta}{2}\right)\right]\exp\left\{ -\frac{n_{*}\theta}{2}\right\}
	=\begin{cases}
	1 & n_{*}\theta\gg1\\
	\frac{\theta n_*}{3} & n_{*}\theta\ll1
	\end{cases}
	\label{f-theta}
	\end{equation}
\end{widetext}
Two terms in the exponent of Eq.(\ref{moments}) come from the averaging over fluctuations of $\xi_{1,2}$\,
(1st term)  and angular variables $\varphi_{1,2}$. 

\textit{OTOC.}
The Out-of-Time Order Correlator is defined as the following irreducible average:
\begin{eqnarray}
\mathcal{F}(\theta_1,\theta_2;\theta_3,\theta_4) =  \langle\langle G(\theta_1,\theta_2)G(\theta_3,\theta_4)\rangle\rangle 
\label{FO1}
\end{eqnarray}
where $\langle ..\rangle$ means averaging w.r.t. fluctuations of $\delta\varepsilon(\varphi)$.
We assume that $ \Re \theta_3>\Re \theta_1> \Re \theta_4 > \Re \theta_2$.
The function $\mathcal{F}$ depends, in general, on 4 independent variables but here we  consider the special case  
$\theta_1=\frac{\pi-\theta}{2}$, $\theta_2= \frac{-\pi-\theta}{2}$, $\theta_3=\frac{\pi+\theta}{2}$, $\theta_4 = \frac{-\pi+\theta}{2}$, see for example Ref.~\cite{Bagrets2017}.
It is convenient  to introduce the function $f(\theta)= \frac{\mathcal{F}(\theta_1,\theta_2;\theta_3,\theta_4) }{G(\theta_1,\theta_2)G(\theta_3,\theta_4)}$;  we calculate its major term of the order of $O(1/N)$ in the limit $N\gg1$. 
After the calculation of $f(\theta)$ within the imaginary time technique, we need to perform analytical continuation
to real times  by the substitution $f(\theta \to -i\varepsilon_0 t) \equiv F(t)$. Exponential growth of $F(t)$
at short times demonstrate quantum-chaotic behaviour of the system.

To calculate $f(\theta)$ we  first write
$G(\theta_1,\theta_2)= \langle G(\theta_1,\theta_2) \rangle + \delta G(\theta_1,\theta_2) $, where
\begin{eqnarray}
\frac{\delta G(\theta_1,\theta_2)}{G(\theta_1,\theta_2)}=\frac{1}{2\pi}\sum_{m\ne 0,\pm 1}(ime^{im\theta_{1}}+ime^{im\theta_{1}}+  \nonumber \\ +\cot\left(\frac{\theta_{1}-\theta_{2}}{2}\right)\left(e^{im\theta_{2}}-e^{im\theta_{1}}\right))\frac{\varepsilon_{m}}{im \varepsilon_0}
\end{eqnarray} 
The same expansion is used for $G(\theta_3,\theta_4)$, and the results are substituted into Eq.(\ref{FO1}), leading
to the following Fourrier series:
\begin{eqnarray}
\label{fo2}
f(\theta) & = &  \frac{1}{(2\pi )^2}\sum_{m\ne 0,\pm 1} 4\cos^2(\frac{m\pi}{2})\cos(m\theta) K_\varepsilon(m) \\ \nonumber
& = & \frac{1}{\pi^2}\sum_{n\neq 0} e^{2 i n\theta}K_\varepsilon(2n)
\end{eqnarray}
where $K_\varepsilon(m)$ is provided by Eq.(\ref{eps-fluct}). 
The series in (\ref{fo2}) can be calculated by the transformation to the integral over $dn$, which is 
determined (after analitic continuation to real $t$ and in the limit $t  \gg 1/T$) 
by the contribution of the single pole of $K_\varepsilon(2n)$ at $n=\frac12$. Finally, we obtain
\begin{eqnarray}
F(t)\approx i\frac{ \varepsilon_0\, e^{2\pi T t} }{2 \varepsilon_0^2 + g(\frac{\pi^2}{4} -2)}
\label{Ff}
\end{eqnarray}
with $\varepsilon_0=2\pi/\tilde\beta \equiv 2\pi \gamma T/J$.
The  Lyapunov exponent $\lambda=2\pi T$ is the same as in the pure SYK$_4$ model, but 
the pre-exponential factor is considerably modified, as second term in denominator of Eq.(\ref{Ff})
dominates at $ T \ll \Gamma$, 
where $F(t) = \frac{i}{N}\frac{T}{T_*}e^{2\pi T t}$ and $T_*=\Gamma^2/J$.


\textit{Conclusions.}  We have shown that a moderate quadratic perturbation to the SYK$_4$ model with $N$ Majorana modes can be described in terms of a self-consistent polaron-type solution.  The presence of such a perturbation with strength $\Gamma$ in the interval $J/N \leq \Gamma \leq J/\sqrt{N}$ stabilizes conformal saddle-point solution for the Majorana Green function within a broad range of energies and temperatures. The SYK$_4$ mean-field Green function $G(\epsilon) \sim 1/\sqrt{J\epsilon}$ (defined at Matsubara frequencies $\epsilon =i\cdot 2\pi T\bigl(n+\frac{1}{2}\bigr)$) is valid down to $T_* \sim \Gamma^2/J$, where a crossover to a Fermi liquid at $\eps\lesssim T_*$ occurs.  Schematic ``phase diagram'' of the model is shown in Fig.1. At low temperatures, there is a genuine phase transition at $\Gamma=\Gamma_c$, where $\Gamma_c$ is defined in (\ref{Gammac}). 
Specifically at $T=0$, the $\epsilon\to +0$ asymptotics changes from $G(\epsilon)\propto \sqrt{\epsilon}$ 
for $\Gamma<\Gamma_c$ to $G(\eps) \simeq 1/\Gamma$ for $\Gamma>\Gamma_c$.
Note that higher-order Green functions $G^{(p)}(\tau)$ display exponential growth with $p$, Eq.(\ref{moments}).
The prefactor of the OTO correlation function (\ref{Ff}) is proportional to $T$ in a low temperature range,
$T_* < T \ll \Gamma $, indicating slighly less chaotic behaviour of the system due to the presence of quadratic perturbations.   These results can be relevant to the description of strongly correlated electron systems with flat bands
of various origin~\cite{mablg1,mablg2,chen18,couto11}.


\textit{Acknowledgments.}
M.V.F. is grateful to L.B.Ioffe, A.Kamenev, V.E.Kravtsov and K.S.Tikhonov for useful discussions. 
Research of A.V.L. was partially supported
by the Foundation for Advancement of Theoretical Physics and Mathematics "Basis" and  Basic research program of HSE. A.K.\ is supported by the Simons Foundation under grant~376205 and through the ``It from Qubit'' program, as well as by the Institute of Quantum Information and Matter, a NSF Frontier center funded in part by the Gordon and Betty Moore Foundation.

\bibliography{paper}

\begin{thebibliography}{25}
\bibitem [{Left estimate follows from the result of of Ref.~\cite{SYKglass},
  which was criticized in Ref.~\cite{SYKglass2}; right estimate corresponds to
  the temperature where entropy of the replica-symmetric solution of SYK$_4$
  model vanishes()}]{glass_comment}%
  \BibitemOpen
  Left estimate follows from the result of of Ref.~\cite{SYKglass}, which was
  criticized in Ref.~\cite{SYKglass2}; right estimate corresponds to the
  temperature where entropy of the replica-symmetric solution of SYK$_4$ model
  vanishes,\ \href@noop {} {}\BibitemShut {NoStop}%
\end{thebibliography}
\newpage

\begin{widetext}
	\begin{center}{\large \bf Supporting Material for "Perturbed Sachdev-Ye-Kitaev model: a polaron in the hyperbolic plane."}
	\end{center}
	\maketitle
	\maketitle
	\section{The effective action}
	In this part, we will describe the solution of the problem using a geometrical approach. The logic will be the same as in the main text. 
	We derive the effective action in adiabatic approximation and then the first non-adiabatic
	correction.  Full action is provided in Eq.(\ref{eq:seff_full}).
	
	\subsection{Adiabatic approximation}
	The action of the SYK model at the Hyperbolic  plane (we use Poincaré disk model) was presented at the main text. After proper regularization it has the form:
	\begin{equation}
	\label{SX}
	S=\int_0^{\tbeta}\left\{\frac{1}{2}g_{\mu\nu} \dot{X}^\mu\dot{X}^\nu-\gamma\omega_{\mu}\dot{X}^\mu \right\} d\tau  - \frac{g\gamma}{4} \int_0^{\tbeta} d\tau_1 d\tau_2  \chi^{1/2}_{z(\tau_1),z(\tau_2)}
	\end{equation} 
	Here $g_{\mu\nu}$ is a metric tensor and $\omega_{\mu}$ is the spin connection.
	We also  introduced the following notations:
	\begin{eqnarray}
	\tbeta=\frac{J \beta}{\gamma},\qquad
	g=\frac{b^{2\Delta}}{2}\frac{N\Gamma^2}{J^2}\gamma^{2-4\Delta}
	= \frac{N\gamma}{4\sqrt{\pi}}\frac{\Gamma^2}{J^2}.
	\label{g}\qquad \chi = \frac{(1-|z_1|^2)(1-|z_2|^2)}{|1-z_1^*z_2|^2}
	\end{eqnarray}
	Here $z$ is a complex coordinate of the point at the model. We will use coordinates $\xi$ and  $\varphi$ which are defined as $z=\tanh(\xi/2)e^{i\varphi}$ to solve our problem.
	We also perform Hubbard–Stratonovich transformation, as a result the action of the problem will be:

	\begin{eqnarray}
	&S_{SYK} = \frac{1}{2}\int_0^{\tbeta}\left[\frac{\dot{\xi}^2}{2}+\sinh^2(\xi)\frac{\dot{\varphi}^2}{2}-\gamma \cosh(\xi)\dot{\varphi}\right] d\tau \nonumber \\
	&S_{\Phi}= \frac{1}{4g\gamma}\int d\mu \Phi(x)(-L-\frac{1}{4}+\delta^2)\Phi(x) \nonumber \\
	&S_{int}=\int_0^{\tbeta} \Phi(x(\tau))d\tau
	\end{eqnarray}

	Here $L$  is the Laplace operator and $d\mu$ is the invariant measure on the hyperbolic plane and we should take a limit $\delta\rightarrow0$.  If we  integrate the bosonic field $\Phi$ we will obtain  the previous action. 
	We employ an adiabatic  approximation,  assuming  that the motion along the phase $\varphi$ is much slower than
	along radial coordinate $\xi$.   Then functional integral  over trajectories  $\xi(\tau)$ can be done
	at fixed value of $\varphi$, which is the way to find  an effective action for $\dot{\varphi}(\tau)$.
	Since parameter $\gamma \gg 1$,  we can use saddle point approximation for $\dot{\varphi}$, which  leads to
	the relation  $\dot{\varphi}=\frac{\gamma\cosh(\xi)}{\sinh^2(\xi)}$. 
	The effective action is then defined in the following way:
	\begin{eqnarray}
	S_{eff} [\varphi(\tau)]
	=\ln\left(\int D\Phi D\xi \delta\left(\dot{\varphi}-\frac{\gamma\cosh(\xi)}{\sinh^2(\xi)}\right)e^{-S}\right)
	\end{eqnarray}
	A Lagrange variable  $\lambda(\tau)$ is used to remove the $\delta$-function.  Then  we need to calculate the  functional integral with the action dependent of trajectories $\xi(\tau)$ and $\lambda(\tau)$:
	\begin{eqnarray}
	\label{S2}
	S & = & S_{\Phi}+S_{int}+\int_0^{\tbeta}
	\left[
	\frac{\dot{\xi}^2}{2} -\frac{1}{2}\gamma^2\frac{\cosh^2(\xi)}{\sinh^2(\xi)} -
	\lambda(\tau) \left( \dot{\varphi}-\frac{\gamma\cosh(\xi)}{\sinh^2(\xi)}\right)
	\right] d\tau \\
	& \simeq & S_{\Phi}+S_{int}+\int_0^{\tbeta}
	\left[
	\frac12\dot{\xi}^2 -  \lambda(\tau) \left( \dot{\varphi}- 2\gamma e^{-\xi(\tau)}\right) 
	\right] d\tau
	- \int_0^\tbeta 2\gamma^2 e^{-2\xi(\tau)} d\tau
	\label{S3}
	\end{eqnarray}
	Representation (\ref{S3}) follows from Eq.(\ref{S2}) since the condition $\gamma \gg 1$ leads also to
	$\xi \gg 1$; we also omit irrelevant constant $\gamma^2/2$.
	Now calculation of the functional integral over $\xi(\tau)$ is reduced  to the solution of the 1D quantum-mechanical 
	problem with the Hamiltonian
	\begin{eqnarray}
	H = -\frac{\partial^2_{\xi}}{2}+2\gamma \lambda(\tau)e^{-\xi}+\Phi(\xi,\varphi(\tau))
	\label{H1}
	\end{eqnarray} 
	It is the same Hamiltonian as one presented in the main text. Its eigenfunctions and eigenvalues will be presented below.
	Last term in the action (\ref{S3}) was neglected in the Hamiltonian (\ref{H1}) due to its smallness w.r.t. other
	terms; however, we will need this term later.  The term $\Phi(\xi,\varphi)$ in Eq.(\ref{H1})  came from 
	$S_{int}$ term in Eq.(\ref{S3}). Explicit form of $\Phi(\xi,\varphi)$ is to be obtained variationally. 
	Variation of the full action over $\Phi$ leads to the relation
	\begin{eqnarray}
	\Phi_{0}(\varphi,\xi)=
	- \int G_{\Phi}(\xi,\varphi|\xi^\prime,\varphi^\prime)\psi_g^2(\xi^\prime,\varphi^\prime)
	\frac{d\varphi^\prime}{\varepsilon(\varphi^\prime)}d\xi^\prime
	\label{Phi01}
	\end{eqnarray}
	where $G_{\Phi}$ is the  Green function of the operator $-L-\frac{1}{4}+\delta^2$, and the limit
	$\delta\rightarrow0$ is implied. Full analysis of this Green function is provided in Sec.IV below;
	here we need its asymptotic expression only (it coincides with Eq.(\ref{Gfinal}) in the end of Sec.IV). 
	$G_{\Phi}(\xi_1,\varphi_1|\xi_2,\varphi_2)=2g\gamma\left(\frac{e^{-\xi_1-\xi_2}}{\varphi^2_{12}}\right)^{1/2}$, 
	where $\varphi_{12}=2 \sin(\frac{\varphi_1-\varphi_2}{2})$.
	
	Using Eq.(\ref{Phi01}) and the result of variation of the full action  over $\lambda(\tau)$, we obtain, as explained
	in the main text:
	\begin{equation}
	\Phi_{0}(\xi,\varphi)= -\frac{\kappa\sqrt{\lambda\gamma}}{2}e^{-\xi/2}\quad \text{where}~ \lambda(\tau)=\frac{\kappa(\kappa-1)}{32\dot{\varphi}}\quad \text{and} \quad \kappa^2 = 32g \ln\left(\frac{\kappa\beta}{16\pi}\right)
	\label{Phi02}
	\end{equation}
	We start our analysis of Eq.(\ref{H1})) from the simplest case of  $\dot{\varphi} = \varepsilon_0 \equiv 2\pi/\tbeta$.
	Then Schrodinger equation (\ref{H1}) with potential (\ref{Phi02}) allows for exact ground-state $\psi_g$ and
	excited  bound-state solutions $\psi_n$.  We provide these functions below together with corresponding eigenvalues,
	assuming $\kappa > 1$:
	\begin{eqnarray}
	\label{psi-g}
	\psi_g(\chi) & = & \frac{e^{-\chi/2}\chi^{\kappa/2-1/2}}{\sqrt{2\Gamma(\kappa-1)}}\, ; \qquad
	E_g = - \frac{(\kappa -1)^2}{32} \\
	\psi_{n}(\chi) & = & \frac{1}{\sqrt{\frac{2\Gamma(n+1)\Gamma(\kappa-n)}{\kappa-2n-1}}}
	e^{-\chi/2}\chi^{(-1-2n+\kappa)/2}U(-n,-2n+\kappa,\chi)\, ; \qquad E_n = - \frac{(1+ 2n - \kappa)^2}{32}
	\label{psi-n}
	\end{eqnarray}
	where $\chi = 8\sqrt{\gamma\lambda} e^{-\xi/2}$ and  $U(n,m,\chi) $ is confluent hypergeometric function;
	line (\ref{psi-n}) is valid for $ 1 + 2n < \kappa $.
	
	Now we need to generalize the above result for non-constant but slowly varying $\dot\varphi \equiv \varepsilon(\varphi)$. 
	Our goal is to determine  effective action $S_{eff}[\varphi(\tau)]$;  equivalent representation can be
	obtained in terms of $S_{eff}[\varepsilon(\varphi)]$, since it is always assumed that 
	$\dot\varphi \equiv \varepsilon(\varphi) > 0$.
	Formally, this functional can be written as
	\begin{eqnarray}
	S_{eff}[\varphi(\tau)]=\left[S_{\Phi}+\int_0^{\beta}E_g(\lambda(\tau),\Phi)d\tau-\int_0^{\beta} \lambda(\tau) 
	\dot{\varphi}d\tau\right]_{saddle}
	\label{eq:seff}
	\end{eqnarray}
	where  "saddle" means that $\Phi$ and $\lambda$ should be determined from the saddle point equations. 

	To find the energy of the ground state for a general choice of $\varepsilon(\varphi)$ it is convenient
	to consider three terms in the Hamiltonian (\ref{H1}) separately and notice that the term which contains
	$\lambda(\varphi)$ is canceled out in the effective action (\ref{eq:seff}). 
	Then we need to calculate the average of the two other terms in the Hamiltonian over the deformed 
	( dependent on $\varepsilon(\varphi)$) ground state:  
	\begin{eqnarray}
	\tilde{E}_g = \frac{\kappa-1}{32} - \int G_{\Phi}(\xi,\varphi|\xi^\prime,\varphi^\prime)\psi_g^2(\xi^\prime,\varphi^\prime)\psi_g^2(\xi,\varphi)\frac{d\varphi^\prime}{\varepsilon(\varphi^\prime)}d\xi^\prime d\xi
	\label{tildeEg}
	\end{eqnarray}
	The first term in (\ref{tildeEg}) comes from kinetic term in the Hamiltonian (\ref{H1}), its dependence on
	$\varepsilon(\varphi)$ is weak and we neglect it in the following. We will estimate its influence below.   The second term, together with 
	$S_{\Phi}$ term in  Eq.(\ref{eq:seff}), combine to our final result for the action in the adiabatic approximation:
	\begin{eqnarray}
	S_{eff} =  -\frac{1}{2} \int G_{\Phi}(\xi,\varphi|\xi^\prime,\varphi^\prime)\psi_g^2(\xi^\prime,\varphi^\prime)\psi_g^2(\xi,\varphi)\frac{d\varphi^\prime d\varphi}{\varepsilon(\varphi^\prime)\varepsilon(\varphi)}d\xi^\prime d\xi = 
	-\frac{g}{2}\int \frac{\kappa-1}{\kappa} \left(\frac{\varepsilon(\varphi_1)\varepsilon(\varphi_2)}{\varphi^2_{12}}\right)^{1/2}\frac{d\varphi_1 d\varphi_2}{\varepsilon(\varphi_1)\varepsilon(\varphi_2)}
	\label{eq:seffII}
	\end{eqnarray}
	For the applicability of our adiabatic approximation strong inequality $\kappa \gg 1$ is needed, 
	thus $\frac{\kappa-1}{\kappa}\approx1$.

	\subsection{Main non-adiabatic correction}
	
	The aim of this Section is to find the first non-adiabatic correction to the action. This correction is due to
	virtual transitions between the levels of the 1D quantum mechanical problem with the Hamiltonian
	(\ref{H1})  which describes motion along coordinate $\xi$. 
	General form of such a correction to $S_{eff}$ is
	\begin{eqnarray}
	\delta S_{eff}=\left[\sum_n \int_{0}^{\beta}d\tau \frac{(\partial_\tau H)_{ng}(\partial_\tau H)_{gn}}{(E_n(\tau)-E_g(\tau))^3}\right]_{saddle}
	\label{Seff-nona}
	\end{eqnarray}
	Here $E_n$ is an energy of the excited state $n$ which adiabatically depends on $\tau$ and $(\partial_\tau H)_{ng}$ is a matrix element of the operator $\partial_\tau H$ between ground state and $n$-th state. 
	Equation (\ref{Seff-nona}) can be obtained applying quantum-mechanical perturbation theory with respect to
	time-dependent terms in the Hamiltonian. The expression (\ref{Seff-nona}) comes in the next order after
	the Berry phase term.
	
	To employ general form (\ref{Seff-nona}) for our purpose,  it is convenient to introduce the following notations:
	\begin{eqnarray}
	M_{n\alpha}=\int_{0}^{\infty}\psi_{n}(\chi)\psi_{g}(\chi)\chi^{\alpha}\frac{2d\chi}{\chi} = \frac{1}{\sqrt{\frac{\Gamma(n+1)\Gamma(\kappa-n)\Gamma(\kappa-1)}{\kappa-2n-1}}}\frac{\Gamma(-1-n+\kappa+\alpha)\Gamma(\alpha+n)}{\Gamma(\alpha)}
	\label{Mn}
	\end{eqnarray} 
	In the limit $\kappa\gg 1$ we have: $M_{n\alpha}=\frac{\Gamma(n+\alpha)}{\Gamma(\alpha)}\kappa^{\alpha-n/2}$.
	Time derivative $\partial H/\partial\tau$ can be written in the form
	\begin{eqnarray}
	\partial_\tau H =2 \gamma \partial_{\tau}\lambda e^{-\xi}-\frac{\kappa\sqrt{\gamma\lambda}\partial_{\tau}\lambda}{4\lambda}e^{-\xi/2}=\frac{\partial_\tau \lambda}{32\lambda}\left(\chi^2-\kappa \chi\right)
	\label{dH}
	\end{eqnarray}
	Using Eq.(\ref{dH}) and notations (\ref{Mn}) we  write:
	\begin{eqnarray}
	(\partial_\tau H)_{gn}=\frac{1}{32}\frac{\partial_\tau \lambda}{\lambda}(M_{n2}-\kappa M_{n1})=\frac{1}{32}\frac{\partial_\tau \lambda}{\lambda}n\kappa^{2-n/2}\sqrt{\Gamma(n+1)}
	\end{eqnarray}
	Here the limit of large $\kappa$ was used to obtain the last result.
	As $E_n =-\frac{1}{32} (-\kappa +2 n+1)^2$  and $\kappa\gg1$ the leading contribution to the $S_{eff}$ comes from the first term in the sum. It brings us to the following expression:
	\begin{eqnarray}
	\delta S_{eff}=\frac{1}{2}\int_0^{\beta} \left(\frac{\partial_\tau \lambda}{\lambda}\right)^2d\tau=\frac{1}{2}\int_0^{2\pi}\frac{d\varphi}{\varepsilon(\varphi)} \left(\partial_{\varphi} \varepsilon(\varphi)\right)^2
	\label{S-n1}
	\end{eqnarray}
	The last expression follows from the expression for $\lambda$ in $(\ref{Phi02})$.
	
	Now we recall the last term in the action (\ref{S3}), which was not taken into account in the adiabatic approximation.
	In the limit of large $\kappa$ the contribution of this term into the ground-state energy can be evaluated as
	$-2\gamma^2\int d\xi \psi_g^2(\xi)e^{-2\xi}$.  Thus  its contribution to the effective action is
	\begin{eqnarray}
	\delta S=-\frac{1}{2}\int_0^{\beta}\int d\xi \psi^2_g(\xi)(2\gamma e^{-\xi})^2\approx-\frac{1}{2}\int_{0}^{2\pi} 
	\frac{d\varphi}{\varepsilon(\varphi)} \varepsilon^2(\varphi)
	\label{S-n2}
	\end{eqnarray}
	Combining the terms in Eqs.(\ref{S-n1},\ref{S-n2}) we find total non-adiabatic contribution to the action
	\begin{eqnarray}
	\delta S_{eff}=-\int_{0}^{\beta} Sch\left\{e^{i\varphi(\tau)},\tau\right\}d\tau 
	\label{eq:seff_cor}
	\end{eqnarray}
	which exactly reproduces the Schwarzian action known for the SYK$_4$ theory.
	Full action is given by the sum of Eq.(\ref{eq:seff_cor}) and Eq.(\ref{eq:seffII}):
	\begin{eqnarray}
	S_{eff}=\frac{1}{2}\int_{0}^{2\pi} \frac{d\varphi}{\varepsilon(\varphi)}\left(\left(\partial_{\varphi} \varepsilon(\varphi)\right)^2-\varepsilon(\varphi)^2\right)-\frac{g}{2}\int \left(\frac{\varepsilon(\varphi_1)\varepsilon(\varphi_2)}{\varphi^2_{12}}\right)^{1/2}\frac{d\varphi_1 d\varphi_2}{\varepsilon(\varphi_1)\varepsilon(\varphi_2)}
	\label{eq:seff_full}
	\end{eqnarray}
	In the next Section we will evaluate fluctuations of $\varepsilon(\varphi)$ controlled by the action (\ref{eq:seff_full}).

	\section{Fluctuation corrections}
	
	In the Section we analyze Gaussian fluctuations of the function  $\varepsilon(\varphi)$ using the action provided in
	Eq.(\ref{eq:seff_full}), and estimate corrections to the fermion Green function related to these fluctuations.
	
	\subsection{Gaussian fluctuations of the $\varepsilon(\varphi)$ function}
	
	Consider the 2nd-order expansion of the action over Fourier-components $\delta \varepsilon_m$ defined as 
	\begin{eqnarray}
	\varepsilon(\theta)=\varepsilon_0+\frac{1}{2\pi}\sum_{m} \delta \varepsilon_m e^{im\theta}
	\end{eqnarray}
	We will assume $\delta\varepsilon(\theta)\ll\varepsilon_0$; equivalently, we  write   $\varphi=\theta+u(\theta)$ and $u(\theta)\ll1$.  Do derive  the action up to quadratic terms in fluctuations, we need to  
	expand $\varepsilon(\varphi)$ up to a second order:
	\begin{eqnarray}
	\varepsilon(\varphi)=\varepsilon_0 \frac{d\varphi}{d\theta}=\varepsilon_0 (1+u^\prime(\theta))\approx \varepsilon_0(1+u^\prime(\varphi)-u(\varphi)u^{\prime\prime}(\varphi))
	\end{eqnarray} 
	The first term in Eq.(\ref{eq:seff_full}) leads to:
	\begin{eqnarray}
	\frac{1}{2}\int_{0}^{2\pi} \frac{d\varphi}{\varepsilon(\varphi)}\left(\left(\partial_{\varphi} \varepsilon(\varphi)\right)^2-\varepsilon(\varphi)^2\right)\approx \frac{\varepsilon_{0}}{2}\int_{0}^{2\pi}d\varphi\left(\left(u^{\prime\prime}\right)^{2}-(1+u^{\prime}u^{\prime})\right) = 
	\frac{1}{4\pi\varepsilon_0}\sum_{m} \delta\varepsilon_{m}\delta\varepsilon_{-m}(m^2-1)
	\end{eqnarray}
	The second term in Eq.(\ref{eq:seff_full}) is not quite trivial to handle,  since the integral over 
	$(\varphi_1-\varphi_2)$ formally diverges, so  some regularization is needed. 
	Explicit regularization with  invariant  short-scale cut-off 
	$\varphi_{12}^2/\varepsilon(\varphi_1)\varepsilon(\varphi_2) > l$ can be used to demonstrate that
	higher harmonics $\varepsilon_m$ are free from this log-divergence.  Since this calculation is relatively
	cumbersome, we present here simpler derivation based on dimensional regularization.
	Namely, we replace power $\frac12$ in the 2-nd term in (\ref{eq:seff_full})
	by some $d < \frac12$ and then take the limit $d \to \frac12-0$.  
	At $d < \frac12$ straightforward Fourier-transformation
	leads to  (with the accuracy up to  terms quadratic in $\varepsilon_m$): 
	\begin{eqnarray}
	\frac{g}{4 \gamma}\int \left(\frac{\varepsilon(\varphi)\varepsilon(\varphi^\prime)}{\varphi_{12}}\right)^{d}\frac{d\varphi^\prime d\varphi}{\varepsilon(\varphi^\prime)\varepsilon(\varphi)} = \frac{1}{2}\frac{g}{4 \gamma} \sum_{m\ne 0} \frac{u_m u_{-m}}{m^2}\int_{0}^{2\pi}\frac{d\varphi}{2\pi} 2(d-1)\varepsilon_0^{2d-4}\left(\frac{1}{4\sin^2(\varphi)}\right)^d ((d-1)\cos(2m\varphi)+d) 
	\label{Intd}
	\end{eqnarray}
	Then  last integral in Eq.(\ref{Intd}) can be calculated using the following formula:
	\begin{eqnarray} 
	\int_0^{2\pi} \frac{d\varphi}{2\pi} \left(\frac{1}{4\sin^2(\varphi)}\right)^d e^{2im\varphi} = \frac{1}{2\cos(\pi d)}  \frac{\Gamma(m+d)}{\Gamma(2d)\Gamma(1+m-d)}
	\label{Intd2}
	\end{eqnarray}
	where $m$ is any integer number. 
	We are interested in the $m$-dependent coefficients which are obtained by derivative of the ratio 
	$\Gamma(m+d)/\Gamma(m+1-d)$ over $d$, evaluated  in the limit $d\to \frac12$. The result reads
	\begin{eqnarray}
	S_{eff}\approx  \frac{1}{4\pi\varepsilon_0}\sum_{m} \delta\varepsilon_{m}\delta\varepsilon_{-m}(m^2-1)+\frac{g}{2}\sum_{m}\frac{\tilde{\psi}(m)}{4\pi\varepsilon_{0}^{3}}\delta\varepsilon_{m}\delta\varepsilon_{-m}
	\label{eq:fluctuation_action}
	\end{eqnarray}
	Here $\tilde{\psi}(x)=\Psi(x+1/2)-\Psi(-1/2)$ and $\Psi(x) = (\ln\Gamma(x))^\prime$ is the digamma function.
	This action leads to the following correlation function:
	\begin{eqnarray}
	\langle \delta\varepsilon_{m}\delta\varepsilon_{-m}\rangle = \frac{2\pi \varepsilon_0^3}{ \varepsilon_{0}^{2}(m^2-1)+\frac{g}{2}\tilde{\psi}(m)}
	\label{correlations-final}
	\end{eqnarray}
	We use it below for calculations of the corrections to fermion Green function.
	
	\subsection{Estimation of the fluctuations of the kinetic term}
	The contribution to the action from the kinetic term has the form:
	\begin{eqnarray}
	S_{kin}=\int\frac{\kappa}{32}d\tau\qquad \kappa^2=32g \ln\left(\frac{\kappa}{8\varepsilon(\varphi)}\right)
	\end{eqnarray}
	Assuming smallness of fluctuations we can write $\kappa=\kappa_0+\delta\kappa$ where $\kappa_0$ is defined by $\varepsilon(\varphi)=\varepsilon_0$. We will also define a parameter $\alpha = \frac{32g}{\kappa_0^2}\ll1$. The connection between $\delta\kappa$ and $\delta\varepsilon$ can be obtained from the definition of $\kappa$ and has the form:
	\begin{eqnarray}
	\delta \kappa =\frac{\kappa_0}{2}\left(\frac{\alpha}{2}\left(\frac{\delta \varepsilon}{\varepsilon_0}\right)^2-\alpha \frac{\delta \varepsilon}{\varepsilon_0}\right)
	\end{eqnarray}	
	This expression leads to the following form of the above action:
	\begin{eqnarray}
	S_{kin}= \frac{1}{2\pi}\frac{g}{2\kappa_0}\sum_{n}\frac{\delta\varepsilon_{n}\delta\varepsilon_{-n}}{\varepsilon_{0}^{2}}
	\end{eqnarray}
	One can see smallness of this part due to the factor $\frac{1}{\kappa \varepsilon_0}\ll1$ with respect to the second term in the $(\ref{eq:fluctuation_action})$

	\subsection{Correction to the Green function}
	Fermion Green function can be obtained as an average 
	of the field $\hat{G}(\theta_1,\theta_2)$, evaluated with the effective action (\ref{eq:seff_full}), where
	\begin{eqnarray}
	\hat{G}(\theta_1,\theta_2)=-\left(b\gamma^2 \frac{\varepsilon(\theta_1)\varepsilon(\theta_2)}{4\sin^2\left(\frac{\varphi(\theta_1)-\varphi(\theta_2)}{2}\right)}\right)^{\Delta}
	\end{eqnarray}
	The saddle point approximation $(\varphi(\theta)=\theta)$ leads to $\langle \hat{G}(\theta_1,\theta_2) \rangle =G_c = -\left(b\gamma^2 \frac{\varepsilon_0^2}{ \theta_{12}^2}\right)^{\Delta}$. 
	We are interested in the quadratic correction to the Green's function. So we need to find the second-order correction  by $\delta\varepsilon$  to $\hat{G}$ :
	\begin{eqnarray}
	&\frac{\delta \hat{G}(\theta_1,\theta_2)}{G_c(\theta_1,\theta_2)}=\frac{1}{2}\sum_{m\ne\pm1,0}\langle\delta \varepsilon_m \delta \varepsilon_{-m}\rangle O_m(\theta_1-\theta_2) \nonumber \\
	&O_m(\theta)=-\frac{\Delta}{(2\pi)^2\sin^{2}\left(\frac{\theta}{2}\right)\varepsilon_{0}^{2}m^{2}}((\Delta(1-m^{2})+1)\cos(m\theta)+\cos(\theta)\left((\Delta-1)m^{2}-\Delta+\Delta\left(m^{2}+1\right)\cos(m\theta)\right) \nonumber \\&-\Delta\left(m^{2}+1\right)+m^{2}+2\Delta m\sin(\theta)\sin(m\theta)-1) \nonumber \\
	\nonumber
	\end{eqnarray} 
	For large $\kappa$ only terms with large $m$ will be important. In this case: $O_m(\theta)=\frac{2\Delta}{(2\pi\varepsilon_0)^2}(\Delta-1+\Delta \cos(m\theta))\sim \frac{2\Delta}{(2\pi \varepsilon_0)^2}$ so we can write
	\begin{eqnarray}
	\frac{\delta \hat{G}(\theta_1,\theta_2)}{G_c(\theta_1,\theta_2)}\sim \frac{1}{2}  \frac{2\Delta}{(2\pi \varepsilon_0)^2}\sum_{m\ne\pm1,0} \langle\delta\varepsilon_m \delta \varepsilon_{-m}\rangle = \frac{1}{2}  \frac{2\Delta}{2\pi }\sum_{m\ne\pm1,0} \frac{ \varepsilon_0}{ \varepsilon_{0}^{2}(m^2-1)+\frac{g}{2}\tilde{\psi}(m)} \sim   \  \frac{\Delta}{\pi } \frac{1}{ \varepsilon_{0}m_*}
	\end{eqnarray}
	Here $m_*$ is defined us $\varepsilon_0^2( m_*^2-1) =\frac{g}{2}\tilde{\psi}(m_*)$. For large $\kappa$ we can write, using Eq.(\ref{Phi02}):  $\varepsilon_0 m_* = \frac{\kappa}{8}$, thus corrections to fermion  Green function are small
	at any $\theta$.
	
	\section{Higher orders of the fermionic Green function.}
	
	The major object of our theory is the Majorana Green function $G(\tau)$  averaged over disorder variables which enter
	the Hamiltonian, Eq.(1) of the main text. However, local Majorana Green function $G_i(\tau,\tau^\prime) = 
	-\langle \chi_i(\tau)\chi_i(\tau^\prime) \rangle $ contains more information about system's dynamics.

	One of the methods to extract this additional information is to consider higher-order Green functions, defined below:
	\begin{equation}
	G^{(p)}(\tau,\tau^\prime) \equiv \langle 
	\left(- \frac1{N}\sum_i \chi_i(\tau)\chi_i(\tau^\prime) \right)^p\rangle
	\label{Gp1}
	\end{equation}
	
	Here we restrict ourselves by the region of moderately high $ p \ll N$, where it is easy to show that
	\begin{equation}
	G^{(p)}(\tau_1,\tau_2)= (-1)^p\left\langle\left[b \frac{e^{-\xi_1-\xi_2}}
	{\sin^{2}(\frac12(\varphi_1-\varphi_2))}\right]^{\Delta p}\right\rangle =
	(-1)^p C^2_{\Delta p} \left\langle\left[\frac{b}{4\gamma} 
	\frac{\varepsilon(\varphi_1)\varepsilon(\varphi_2)}
	{\sin^{2}(\frac12(\varphi_1-\varphi_2))}\right]^{\Delta p}\right\rangle_{S_\varphi}
	\label{Gp2}
	\end{equation}
	Angular brackets in the middle formula of the above equation mean averaging over quantum action $S_{eff}$,
	see Eq.(11) of the main text. Formula in the R.H.S. of (\ref{Gp2}) is obtained after we take average over
	fluctuations of $\xi_1$ and $\xi_2$ over the polaron ground state $\psi_g(\xi)$, where $C_\alpha$ is defined below:
	\begin{equation}
	C_{\alpha}=\left(\frac{2\gamma}{\varepsilon(\varphi)}\right)^{\alpha}\int e^{-\alpha\xi}\psi_{g}^{2}(\xi,\varphi)d\xi=\frac{\Gamma(\kappa+2\alpha-1)}{\Gamma(\kappa-1)}\kappa^{-\alpha}(\kappa-1)^{-\alpha}  \approx 
	\exp\left(\frac{2\alpha^2}{\kappa}\right)
	\label{Ca}
	\end{equation}
	We used assumption $\alpha\ll\kappa$ to make the last approximation.
	Final averaging over $S_\varphi$ in the R.H.S. of Eq.(\ref{Gp2}) should be done with the full phase-dependent 
	action given by  Eq.(\ref{eq:seff_full}). Last expression in Eq.(\ref{Ca}) is valid in the main order of approximation
	for $\kappa \gg 1$ and $\alpha \gg 1$.
	
	Consider now the effect of integration over fluctuations of angular modes $\varepsilon(\varphi)$ and define 
	relevant measure for these fluctuations
	\begin{eqnarray}
	g_{p}(\tau_{1},\tau_{2})=\frac{\langle G^{(p)}(\tau_{1},\tau_{2})\rangle}{C_{\Delta p}^{2}G_{c}^{p}(\tau_{1},\tau_{2})}
	=\langle\exp\left[\Delta p\delta g(\theta_{1},\theta_{2})\right]\rangle =
	\exp\left(\frac{(\Delta p)^2}{2} \langle (\delta g(\theta_{1},\theta_{2}))^2 \rangle \right)
	\label{gp}
	\end{eqnarray}	
	where $G_c(\tau_1,\tau_2)$ is the conformal saddle-point Green function, while the function $\delta g(\theta_{1},\theta_{2})$
	is defined via the relation
	\begin{eqnarray}
	\frac{\varepsilon(\varphi_{1})\varepsilon(\varphi_{2})}{4\sin^{2}(\frac{\varphi_{1}-\varphi_{2}}{2})}\cdot
	\left[\frac{\varepsilon_{0}\varepsilon_{0}}{4\sin^{2}(\frac{\theta_{1}-\theta_{2}}{2})}\right]^{-1}
	\equiv 1+ \delta g(\theta_{1},\theta_{2}) = 1+
	u^{\prime}(\theta_{1})+u^{\prime}(\theta_{2})+\cot\left(\frac{\theta_{1}-\theta_{2}}{2}\right)\left(u(\theta_{2})-u(\theta_{1})\right)
	\label{gp2}
	\end{eqnarray}	
	We use here definitions $\varphi=\theta+u(\theta)$ and $\varepsilon(\varphi)=\varepsilon_0\frac{d\varphi}{d\theta}$.
	To calculate the average in the R.H.S. of Eq.(\ref{gp}) we need to expand the R.H.S. of Eq.(\ref{gp2}) up
	to linear terms in $u(\theta)$ and then use  Fourier series:
	\begin{equation}
	\delta g(\theta_{1},\theta_{2}) =
	\frac{1}{2\pi}\sum_{m}\left(ime^{im\theta_{1}}+ime^{im\theta_{1}}+\cot\left(\frac{\theta_{1}-\theta_{2}}{2}\right)\left(e^{im\theta_{2}}-e^{im\theta_{1}}\right)\right)u_{m}
	\label{gp3}
	\end{equation}
	Now we can average  R.H.S. of Eq.(\ref{gp}) in the Gaussian approximation,
	using representation (\ref{gp3}) and correlation function defined in (\ref{correlations-final}).
	Correlation function in the $\theta$-representation is (below $\theta=\theta_1-\theta_2$):
	\begin{eqnarray}
	\label{gp4}
	\langle\delta g^{2}(\theta_{1},\theta_{2})\rangle & = & \frac{1}{(2\pi)^{2}}\sum_{m}\left(2m\cos\left(\frac{m\theta}{2}\right)-2\cot\left(\frac{\theta}{2}\right)\sin\left(\frac{m\theta}{2}\right)\right)^{2}\langle u_{m}u_{-m}\rangle
	\\ \nonumber
	& \approx & 
	\frac{1}{2\pi\varepsilon_{0}}\Re\sum_{m\ne0,\pm1}\frac{1}{m^{2}}\frac{1}{m^{2}+m_{*}^{2}}\left[2m^{2}\left(1+e^{im\theta}\right)+4im\cot\left(\frac{\theta}{2}\right)e^{im\theta}+2\cot^{2}\left(\frac{\theta}{2}\right)\left(1-e^{im\theta}\right)\right]  \\  \nonumber
	& = & \frac{1}{\varepsilon_{0}}\frac{\left(2+m_{*}\theta\right)}{m_{*}^{3}\theta^{2}}\left[2m_{*}\theta\cosh\left(\frac{m_{*}\theta}{2}\right)-4\sinh\left(\frac{m_{*}\theta}{2}\right)\right]\exp\left\{ -\frac{m_{*}\theta}{2}\right\} 
	\equiv \frac{8}{\kappa}f(\theta)
	\end{eqnarray}	
	where  $\varepsilon_0m_* = \kappa/8$ and last equality just defines a convenient notation. 
	Asymptotic limits for the function $f(\theta)$ are given by
	\begin{eqnarray}
	f(\theta)=\begin{cases}
	1 & m_{*}\theta\gg1\\
	\frac{\theta m_*}{3} & m_{*}\theta\ll1
	\end{cases}
	\end{eqnarray}
	Finally, combining Eqs.(\ref{Gp2},\ref{Ca},\ref{gp},\ref{gp4}) and replacing $\Delta \to \frac14$ we obtain
	\begin{equation}
	\frac{G^{(p)}(\tau_1,\tau_2)}{[G(\tau_1,\tau_2)]^p} = 
	\exp\left[\frac{p^2}{4\kappa}\left(1 + f(\theta_{12})\right)\right]
	\label{Gpf}
	\end{equation}

	\section{The Green function of the boson field on the hyperbolic plane.}
	
	The action of the bosonic field is
	\begin{eqnarray}
	S_{\Phi}= \frac{1}{2g}\int d\mu \Phi(x)(-L-\frac{1}{4}+\delta^2)\Phi(x)
	\end{eqnarray}
	Here $L$ is the Laplace operator and $d\mu$ is an invariant measure on the hyperbolic plane and $\delta\rightarrow0$. We use the Poincaré disk model.
	The Green function of the bosonic field satisfy the following equation:
	\begin{eqnarray}
	(-L-\frac{1}{4}+\delta^2)G(z_1,z_0)=g\frac{\delta(z_1-z_0)}{\sqrt{g(x_0)}}
	\end{eqnarray}
	All objects here are invariant under $SL(2,R)$ transformations so let us use transforms which maps $z_0\mapsto0$ in this case $z_1 \mapsto \frac{z_1-z_0}{1-z_1 \bar{z}_0}$. In new coordinates the form of equation will be the same but $\delta$ function will be localized in the origin of the hyperbolic plane so we expect the rotation invariant solution. It leads us to the equation:
	\begin{eqnarray}
	\left[-(1-u)^2(u\partial^2_u+\partial_u)-\frac{1}{4}+\delta^2\right]G(z)=g\frac{\delta(u)}{4\pi}
	\end{eqnarray}
	Here $u=|z|^2$. This equation can be written as the homogeneous equation with boundary conditions: the Green 
	function should decay faster than $(1-u)^{1/2}$ at $ u \to 1$, while at $u \ll 1$ it should behave as 
	$G(u)\rightarrow -\frac{\ln(u)}{4\pi}$. Then we come  to the following result:
	\begin{eqnarray}
	G(u)=g\frac{1}{4}(1-u)^{\frac{1}{2}+\delta}{}_2F_1\left(\frac{1}{2}+\delta,\frac{1}{2}+\delta,1+2\delta,1-u\right)
	\end{eqnarray}
	Here ${}_2F_1(a,b,c;x)$ is a hypergeometric function.
	In the limit $\delta\rightarrow0$ 
	\begin{eqnarray}
	G(z_1,z_0)=g\frac{\sqrt{w}K(w)}{2\pi} \quad \text{where} \quad 
	w=\frac{(1-|z_1|^2)(1-|z_0|^2)}{(1-z_1\bar{z}_0)(1-z_0\bar{z}_1)}
	\end{eqnarray}
	Here $K(w)$ is the complete elliptic integral of the first kind.
	In the limit $w\rightarrow0$ we have:
	\begin{eqnarray}
	G_\Phi(z_1,z_0)\approx \frac{g}{4}w^{1/2}
	\label{Gfinal}
	\end{eqnarray}
	It is the last form (\ref{Gfinal}) for the Bose field Green function $G_\Phi$,
	which we use in the main text and in Sec.I above.
\end{widetext}
\end{document}